\documentclass[12pt]{iopart}
\usepackage{graphicx,iopams}
\usepackage{citesort}
\usepackage{color}
\usepackage{lineno}
\usepackage{amssymb}
\usepackage{epstopdf}

\begin{document}

\title{The anti-Fermi-Pasta-Ulam-Tsingou problem in one-dimensional diatomic lattices}
\author{Sihan Feng$^{1}$,~Weicheng Fu$^{2,3*}$,~Yong Zhang$^{1,3\dag}$,~Hong Zhao$^{1,3}$}

\address{$^1$Department of Physics, Xiamen University, Xiamen 361005, Fujian, China\\
$^2$Department of Physics, Tianshui Normal University, Tianshui 741001, Gansu, China\\
$^3$Lanzhou Center for Theoretical Physics, Key Laboratory of Theoretical Physics of Gansu Province, Lanzhou University, Lanzhou, Gansu 730000, China
}
\ead{$^{*}$fuweicheng@tsnu.edu.cn;~$^{\dag}$yzhang75@xmu.edu.cn}

\date{\today}

\begin{abstract}
We study the thermalization dynamics of one-dimensional diatomic lattices (which represents the simplest system possessing multi-branch phonons), exemplified by the famous Fermi-Pasta-Ulam-Tsingou (FPUT)-$\beta$ and the Toda models. Here we focus on how the system relaxes to the equilibrium state when part of highest-frequency optical modes are initially excited, which is called the anti-FPUT problem comparing with the original FPUT problem (low frequency excitations of the monatomic lattice). It is shown numerically that the final thermalization time $T_{\rm eq}$ of the diatomic FPUT-$\beta$ chain depends on whether its acoustic modes are thermalized, whereas the $T_{\rm eq}$ of the diatomic Toda chain depends on the optical ones; in addition, the metastable state of both models have different energy distributions and lifetimes. Despite these differences, in the near-integrable region, the $T_{\rm eq}$ of both models still follows the same scaling law, i.e., $T_{\rm eq}$ is inversely proportional to the square of the perturbation strength. Finally, comparisons of the thermalization behavior between different models under various initial conditions are briefly summarized.
\end{abstract}

\section{\label{sec:1}Introduction}

Since Fermi, Pasta, Ulam, and Tsingou (FPUT) studied the thermalization problem in one-dimensional (1D) nonlinear oscillator chains in the 1950s \cite{Fermi1955,dauxois2008PhyToday}, this topic has aroused extensive research interest and stimulated many research fields \cite{Campbell2005Chaos,Zabusky2005Chaos,Zaslavsky2005Chaos,Berman2005Chaos,Carati2005Chaos,2008LNP728G} (see also references therein). Thenceforth 1D nonlinear oscillator chains become the testbed for exploring the dynamics and the statistical properties of many-body interactions. The core of the studies on the subject lies in how a nonlinear mechanical system relaxes to the thermalized state from the various initial conditions far from equilibrium. Following the seminal work of FPUT where the energy is initially injected into one or a few lowest-frequency modes, it is found that solitons \cite{ZabuskySoliton} or $q$-breathers \cite{PhysRevLett.95.064102,FlachPRE2006} will form in the system at the early stage of relaxation, which blocks the process of thermalization. However, when the highest-frequency modes are initially excited, which can be traced back to Zabusky's work in 1967 \cite{ZABUSKY1967126}, completely different dynamics are observed. It is found that the system will form chaotic breathers \cite{CRETEGNY1998109,Zabusky2006Chaos}, which also delay the system from entering the thermalized state. To distinguish from the original work of FPUT where the initial lowest-frequency excitations (LFE) were considered, the studies under the initial highest-frequency excitations (HFE) are named, first by Dauxois et al. \cite{dauxois2005anti}, as the anti-FPUT problem. Although the dynamical behavior of the initial stage presents differences between the FPUT and anti-FPUT cases, the dependence of the thermalization time $T_{\rm eq}$ on the energy density follows similar power laws \cite{PhysRevE.61.2471,MIRNOV2001251}. Recent studies have shown that these scaling laws can be explained in the framework of wave turbulence theory \cite{Onorato4208,PhysRevLett.120.144301,Pistone2018EPL}. Subsequently, it is shown that, in the thermodynamic limit, the thermalization behavior of a near-integrable system is universal as long as the ability of the system to be thermalized is properly measured, that is, the $T_{\rm eq}$ is inversely proportional to the square of the perturbation strength \cite{PhysRevE.100.010101,Fu_2019Toda,PhysRevE.94.062104,PhysRevE.100.052102,Pistone2018,PhysRevLett.124.186401}. The key to accurately describing the thermalization capability of a system is to define the perturbation strength by selecting a suitable reference integrable system \cite{Fu_2019Toda}.

The diatomic chain, with periodically arranged atoms of two different masses, is another typical model in the sense that isotopic mass impurities do occur in nature. The introduction of unequal masses into the lattice leads to some important new properties. For example, the phonons of the diatomic chains are polarized into acoustic and optical branches. Such polarization may bring about the three-wave interaction which however is forbidden in the monatomic chains \cite{pezzi2021threewave,Onorato4208}. Analogously to the monatomic chains, early studies have also found solitons \cite{collins1985solitons,vainchtein2016solitary} and breathers \cite{zhou1996comparative} in the diatomic cases. Yet, to the best of our knowledge, there is little research on how such a property affects the thermalization dynamics of diatomic chains. Recently, we have studied the FPUT problem in the diatomic FPUT-$\beta$ and Toda chains, where the initial energy is fed into the lowest-frequency acoustic modes, and found that the thermalization of the system also follows the inverse square law although the integrability of the two systems is broken in different ways \cite{PhysRevE.100.052102}.

In the present work, we focus on how a diatomic lattice relaxes to the thermalized state when part of the highest-frequency optical modes are initially excited, i.e., the anti-FPUT problem of diatomic lattices. As illustrating examples, the diatomic FPUT-$\beta$ chains and the diatomic Toda chains are studied systemically. The rest of the paper is structured as follows. The models are introduced in Sec.~\ref{sec:2}. The physical quantities and numerical methods are presented in Sec.~\ref{sec:3}, followed by the numerical results in Sec.~\ref{sec:4}. And finally, the concluding remarks are given in Sec.~\ref{sec:5}.

\section{\label{sec:2}The models}

 \begin{figure}[ht]
  \centering
  \includegraphics[width=1\columnwidth]{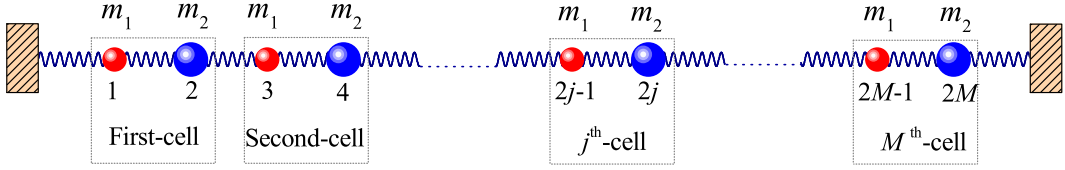}\\
  \caption{Schematic of the 1D diatomic chain with $M$ unit cells.}\label{figDiatomicChain}
 \end{figure}
We consider a 1D diatomic lattice consisting of $M$ unit cells with fixed ends as shown in Fig.~\ref{figDiatomicChain}. Each unit cell contains two particles of mass $m_1$ and $m_2$ situated alternately at the position $2j-1$ and $2j$ in the $j$th unit cell, and the total number of particles is $N=2M$; its Hamiltonian is
 \begin{equation}\label{eqHariltonian}
  H=\sum_{i=1}^{N}\left[\frac{p_i^2}{2m_i}+V(q_i-q_{i-1})\right],
 \end{equation}
where $m_i$, $p_i$, and $q_i$ are, respectively, the mass, momentum, and displacement from the equilibrium position of the $i$th particle, and $V$ is the nearest-neighboring interparticle interaction potential. Without losing generality, we set $m_1=1-\Delta m/2$, and $m_2=1+\Delta m/2$ ($|\Delta m|<2$ to guarantee positive masses, and $\Delta m>0$ considered mainly) such that the mass density is fixed to be unity. To facilitate the analysis in the framework of perturbation theory, the Hamiltonian of a system is usually written as
 \begin{equation}
  H=H_0+H',
 \end{equation}
where $H_0$ denotes the integrable part, and $H'$ is the perturbation. The form and the strength of $H'$ completely depend on the choice of $H_0$. And the existing results show that the choice of $H_0$ is very important to characterize the thermalization behavior of a system \cite{Fu_2019Toda,PhysRevE.100.052102,PhysRevE.104.L032104}.

In this work, we consider two kinds of interaction potentials. The first one is the FPUT-$\beta$ which reads
 \begin{equation}
  V_{\beta}(x)=\frac{x^2}{2}+\frac{\beta  x^4}{4},
 \end{equation}
where $\beta$ is a free and positive parameter. The FPUT-$\beta$ potential is symmetric since $V_{\beta}(x) = V_{\beta}(-x)$. When $\beta=0$, it becomes a harmonic (integrable) one, so here the Hamiltonian of a diatomic harmonic chain is taken as $H_0$, namely,
\begin{equation}\label{HDharm}
  H_0=\sum_{i=1}^{N}\left[\frac{p_i^2}{2m_i}+\frac{1}{2}(q_i-q_{i-1})^2\right],
\end{equation}
then the fourth-order nonlinearity is the perturbation which has the form of
\begin{equation}\label{epsilonBeta0}
  H'=H-H_0=\sum_{i=1}^{N}\left[\frac{\beta}{4}(q_i-q_{i-1})^4\right].
\end{equation}
Via rescaling the relative displacement with the energy density $\varepsilon$ (i.e., the energy per particle) so that $\tilde{q}=q/\sqrt{\varepsilon}$, we can obtain the dimensionless perturbation strength as
\begin{equation}\label{epsilonBeta}
  \widetilde{H}'=\frac{H'}{\varepsilon}=\frac{\beta\varepsilon}{4}\sum_{i=1}^{N}\left[(\tilde{q}_i-\tilde{q}_{i-1})^4\right]\sim \beta \varepsilon.
\end{equation}
Therefore, it is equivalent to study the dynamical behavior of the system via varying $\beta$ by fixing $\varepsilon$ or that of varying $\varepsilon$ by fixing $\beta$. The second one is the Toda potential \cite{1967Toda} which takes the form as
\begin{equation}
 \label{eq:Toda}
  V_{\rm T}(x)=\frac{e^{2 x}-2 x-1}{4},
\end{equation}
which is an asymmetric one since $V_{\rm T}(x)\neq V_{\rm T}(-x)$. It is shown that, for a given $\varepsilon$, the diatomic Toda chain can be regarded as the perturbed Toda by unequal masses. It is easily to prove that, from the Hamilton canonical equation, the dynamical system described by Eq.~(\ref{eqHariltonian}) is strictly equivalent to the homogeneous chain with unit mass depicted by the following Hamiltonian:
\begin{equation}\label{eqHariltonian2}
  H=\sum_{i=1}^{N}\left[\frac{v_i^2}{2}+\frac{1}{m_i}V(q_i-q_{i-1})\right],
\end{equation}
where $v_i$ is the velocity of $i$th particle, and $1/m_i$ is the renormalization coefficient of the force constant related to the lattice site. Then we take the Hamiltonian of the Toda chain as $H_0$ with the form of
\begin{equation}\label{eqHToda}
  H_0=\sum_{i=1}^{N}\left[\frac{v_i^2}{2}+\frac{1}{m_1}V_{\rm T}(q_i-q_{i-1})\right],
\end{equation}
thus, with the help of Eqs.~(\ref{eqHariltonian2}) and (\ref{eqHToda}), we can obtain the perturbation as
\begin{equation}\label{eqHpToda}
  H'=H-H_0=\frac{|m_2-m_1|}{m_1m_2}\sum_{i=2,4,6,\dots}^{N}V_{\rm T}(q_i-q_{i-1}),
\end{equation}
and the average perturbation strength for a given energy density
\begin{equation}\label{ptbDeltM}
   \langle H'\rangle= \frac{|m_2-m_1|}{m_1m_2}\left\langle\sum_{i=2,4,6,\dots}^{N}V_{\rm T}(q_i-q_{i-1})\right\rangle\sim\frac{\Delta m}{1-\Delta m^2/4}\sim\Delta m,
\end{equation}
for small $\Delta m$ (since we here focus on the thermalization properties in the near-integrable region), where $\langle \cdot\rangle$ represents the ensemble average of the thermodynamic equilibrium state. From another perspective, when the $\Delta m$ is fixed, the variation of $\varepsilon$ will also change the strength of nonintegrability of the diatomic Toda chain. To see this clearly, Eq.~({\ref{eq:Toda}) can be expanded by the Taylor series at the zero point of the potential (i.e., $x = 0$) as follows
\begin{equation}\label{eq:TodaExpand}
  V_{\rm T}(x)=\frac{x^2}{2}+\frac{x^3}{3}+\sum_{n=4}^{\infty}\frac{2^{n-2}x^n}{n!}.
\end{equation}
A given nonzero $\Delta m$ breaks the Toda's integrability \cite{casati1975stochastic}, such that the 1D diatomic Toda chain can be regarded as the diatomic harmonic one, i.e., Eq.~(\ref{HDharm}) plus the high-order nonlinearity perturbation, and the dimensionless perturbation
\begin{equation}\label{eq:TodaExpande0}
   \widetilde{H}'=\frac{H'}{\varepsilon}=\frac{\varepsilon^{1/2}}{3}\sum_{i=1}^{N}(\tilde{q}_i-\tilde{q}_{i-1})^3+
\sum_{n=4}^{\infty}\frac{2^{n-2}\varepsilon^{(n/2-1)}}{n!}\sum_{i=1}^{N}(\tilde{q}_i-\tilde{q}_{i-1})^n,
\end{equation}
and thus the leading perturbation strength
\begin{equation}\label{eq:TodaExpande}
   \widetilde{H}'\sim \varepsilon^{1/2}
\end{equation}
for small energy density $\varepsilon$.

\section{\label{sec:3}Physical quantities and numerical method}

The orthogonalized, normalized eigenvector, $\mathbf{U}_k$, of a 1D diatomic chain for the fixed ends can be given as
\begin{eqnarray}
  \mathbf{U}_{k}^{+}=\frac{\mathbf{u}_k^{+}}{\|\mathbf{u}_k^{+}\|}, \label{eq1}\\
  \mathbf{U}_k^{-}=\frac{\mathbf{\widetilde{U}}_k^{-}}{||\mathbf{\widetilde{U}}_k^{-}\|},~\mathrm{and}~ \mathbf{\widetilde{U}}_k^{-}=\mathbf{u}_k^{-}-\frac{\langle \mathbf{u}_k^{-},\mathbf{U}_k^{+}\rangle}{\| \mathbf{U}_k^{+}\|^2}\mathbf{U}_k^{+},
  \label{eq2}
\end{eqnarray}
where symbols `$-$' and `$+$', respectively, correspond to the acoustic branch and the optical branch, and $k=1,2,\dots,M$. The symbol $\|\cdot\|$ represents the length of a vector, and
$\langle \mathbf{A},\mathbf{B}\rangle$ denotes the inner product of vectors $\mathbf{A}$ and $\mathbf{B}$. The element of $\mathbf{u}_k^{\pm}$ is
 \begin{equation}\label{eq:Vec}
 \label{cases}
  u_{i,k}^{\pm}=\cases{
    \frac{\sin\left(\frac{2jk\pi}{2M+1}\right)+
    \sin\left(\frac{2(j-1)k\pi}{2M+1}\right)}{\sin\left(\frac{2k\pi}{2M+1}\right)},&for $i=2j-1$;\\
    \frac{\left[2-m_1\left(\omega_k^{\pm}\right)^2\right]\sin\left(\frac{2jk\pi}{2M+1}\right)}{\sin\left(\frac{2k\pi}{2M+1}\right)},&for $i=2j$,
 }
 \end{equation}
where $\omega_k^{\pm}$ is the frequency of the $k$th eigenvector, as shown below:
\begin{equation}
  \omega_k^{\pm}=\sqrt{\frac{m_1+m_2}{m_1m_2}\left[1\pm\sqrt{1-\frac{4m_1m_2}{(m_1+m_2)^2}\sin^2\left(\frac{k\pi}{2M+1}\right)}\right]}.
\end{equation}
To facilitate the description of the following results, $N$ frequencies follow an ascending order through removing the symbols `$\pm$', i.e, $\omega_{k}=\omega_{k}^{-}$, and $\omega_{N-k+1}=\omega_{k}^{+}$ for $1\le k\le M$. The superscript `$\pm$' of the corresponding eigenvector is also removed, i.e., $\mathbf{U}_{k}=\mathbf{U}_{k}^{-}$, and $\mathbf{U}_{N-k+1}=\mathbf{U}_{k}^{+}$ (see Ref. \cite{PhysRevE.100.052102} for details). Consequently, the normal modes of the 1D diatomic lattice are defined as
\begin{equation}
  Q_k(t)=\sum_{i=1}^{N}q_i(t)U_{i,k},\quad P_k(t)=\sum_{i=1}^{N}p_i(t)/m_iU_{i,k},\quad k=1,2,\dots,N.
\end{equation}
The energy of the $k$th normal mode is
 \begin{equation}
  E_k(t)=\frac{1}{2}\left[P_k^2(t)+\omega_k^2Q_k^2(t)\right].
 \end{equation}
The amplitude $Q_k$, canonical momentum $P_k$, and energy $E_k$ of the $k$th normal mode satisfy the following relationship
\begin{equation}\label{eqPhk}
  Q_k(t)=\sqrt{2E_k(t)/\omega_k^2}\sin{\left(\varphi_k\right)},\quad P_k(t)=\sqrt{2E_k(t)}\cos{\left(\varphi_k\right)},
\end{equation}
where $\varphi_k$ is the phase of the mode. Following the definition of equipartition, it is expected that
 \begin{equation}
  \lim_{T\rightarrow\infty}\bar{E}_k(T)\simeq\varepsilon, \quad k=1,~\dots,~N,
 \end{equation}
where $\bar{E}_k(T)$ denotes the time average of $E_k$ up to time $T$,
 \begin{equation}\label{eq:EkT}
  \bar{E}_k(T)=\frac{1}{(1-\mu)T}\int_{\mu T}^TE_k(t)dt,
 \end{equation}
where $\mu\in[0,1)$ controls the size of the window of time average, and $\mu=2/3$ is fixed throughout this work.

Based on the above preparations, we can introduce the normalized effective relative number of degrees of freedom~\cite{PhysRevA.31.1039},
 \begin{equation}\label{eq:xi}
  \xi(t)=N^{-1}e^{\eta(t)},
 \end{equation}
to measure how close the system is to the state of equipartition, where
\begin{equation}
  \eta(t)=-\sum_{k=1}^{N}w_k(t)\log[w_k(t)]
\end{equation}
is the spectral entropy and
\begin{equation}
  w_k(t)=\bar{E}_k(t)/\sum_{j=1}^{N}\bar{E}_j(t).
\end{equation}
When the system enters the state of equipartition, $\xi(t)$ will saturate at the value $1$.

In our molecular dynamics simulation, the equations of motion are numerically integrated by the eighth-order Yoshida algorithm~\cite{YOSHIDA1990262} with a typical time step $\Delta t=0.1$, and the corresponding relative energy conservation error is less than $10^{-5}$. To suppress fluctuations, all numerical results shown below are the ensemble average denoted by $\langle \cdot \rangle$ that is done over $120$ different random choices of the phases uniformly distributed in $[0,2\pi]$, i.e., in Eq.~(\ref{eqPhk}), $\phi_k\in[0,2\pi]$ is a random number and the initial energy $E_k(0)$ of the $k$th excited mode is a positive random value under the constraint that $\sum_k E_k(0)=N\varepsilon$ and the summation is done over the region of excited modes. In all the calculations below, unless otherwise noted, the highest $10\%$ of the frequency modes are initially excited. We have verified that no qualitative difference will result when the percentage of the initially excited modes is changed.

\section{\label{sec:4}Numerical results}

 \begin{figure}[t]
  \centering
  \includegraphics[width=1\columnwidth]{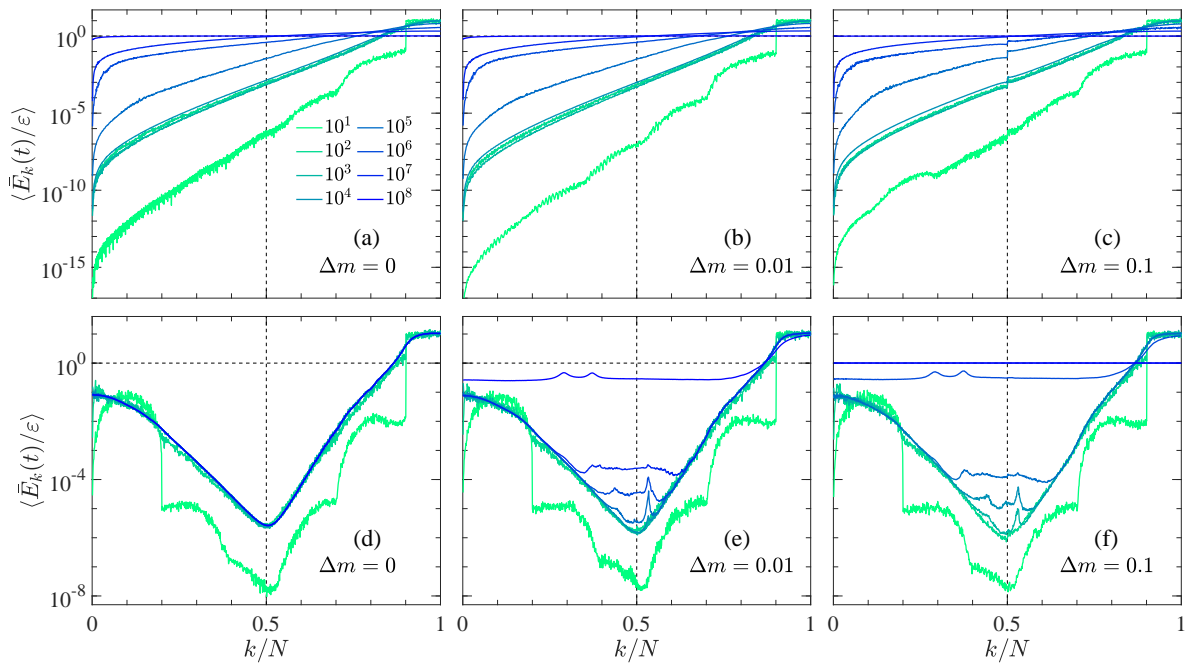}\\
  \caption{(a)-(c) The function of $\langle \bar{E}_k(t)/\varepsilon\rangle$ against $k/N$ at the selected times [the different color lines are for various time $T=10^1, \dots, 10^8$, see the legend, which is applied for all panels, in panel (a)] for the diatomic FPUT-$\beta$ chain with various mass difference $\Delta m$, in the semilogarithmic scale. (d)-(f) The results for the diatomic Toda chain. The number of particles~$N=1024$, and the energy density $\varepsilon=0.01$ are kept fixed.}\label{fig:1}
 \end{figure}

Figures~\ref{fig:1}(a)$-$\ref{fig:1}(c) show the results of $\langle \bar{E}_k(t)/\varepsilon\rangle$ versus $k/N$ at selected times for the diatomic FPUT-$\beta$ chains with different $\Delta m$. Comparing the curves of energy distribution at the same time in the three panels, it is seen that the shapes are almost identical, which means that the $\Delta m$ does not affect the thermalization behavior of the diatomic FPUT-$\beta$ chain, since the variation of $\Delta m$ only changes $H_0$ but not the perturbation strength (see Ref.~\cite{PhysRevE.100.052102} for quantitative analysis). In detail, it can be seen that the energy assigned initially to the highest-frequency optical modes gradually transports to the lowest-frequency acoustic ones and forms an exponential distribution with the evolution of the system. Figures~\ref{fig:1}(d)$-$\ref{fig:1}(f) are numerical results for the diatomic Toda chains. From Fig.~\ref{fig:1}(d), it can be seen that, when $\Delta m=0$, the energy initially injected into the highest-frequency optical modes rapidly transfers to the lowest-frequency acoustic modes first, and then quickly forms a stable V-shaped distribution unchanging over time. It indicates that the thermalized state can never be reached due to the integrability of the monatomic Toda chain. When $\Delta m \neq 0$, from Figs.~\ref{fig:1}(e) and \ref{fig:1}(f), it is seen that the energy distribution of the diatomic Toda chain rapidly reaches the V-shaped curve of the monatomic Toda one first, then the modes near the boundary of Brillouin zone (i.e., the vertical dashed line) obtain energy, and the system eventually enters the thermalized state [see Fig.~\ref{fig:1}(f)]. It means that the early thermalization dynamics of diatomic Toda chains still show characteristics of the integrability of the monatomic Toda chains although it has been broken by unequal masses. Note that, for the diatomic Toda chain, there is a qualitative difference between the V-shaped energy distribution formed under HFE and the exponential distribution formed under LFE (see Fig.~3 in Ref.~\cite{PhysRevE.100.052102}). Indeed this difference is led by the selection rule of modes in an asymmetric interaction model \cite{BIVINS197365,SHOLL1990253}. In addition, comparing the results of the two models, it can be seen from Fig.~\ref{fig:1} that the highest-frequency modes of diatomic Toda chains are difficult to be thermalized, while the lowest-frequency modes of diatomic FPUT-$\beta$ chains are difficult to be thermalized.

\begin{figure}[t]
  \centering
  \includegraphics[width=1\columnwidth]{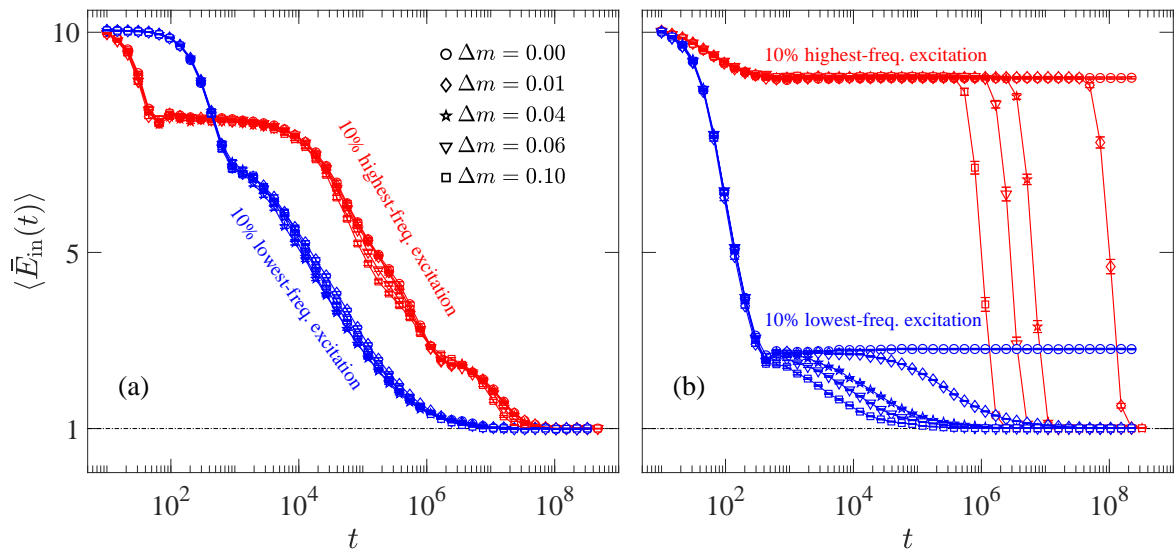}\\
  \caption{(a) The function of energy $\langle\bar{E}_{\rm in}(t)\rangle$ versus time $t$ for the diatomic FPUT-$\beta$ chain with different initial conditions of excitation, and the different $\Delta m$, in semilogarithmic scale. (b) The results for the diatomic Toda chain with the same conditions in panel (a). The number of particles~$N=1024$, and the energy density $\varepsilon=0.01$ are kept fixed.}\label{fig:EkvsT}
\end{figure}

To quantitatively characterize the relaxation behavior of initial energy in different models under conditions of HFE and LFE, we track the evolution of the average normalized energy of the modes initially excited, namely,
\begin{equation}
  \bar{E}_{\rm in}(t)=\frac{1}{\theta_{\rm 2} - \theta_{\rm 1}}\sum_{k=\theta_{\rm 1}+1}^{\theta_{\rm 2}}\bar{E}_k(t)/\varepsilon,
\end{equation}
where $\theta_{\rm 1} = 0$ and $\theta_{\rm 2} = 0.1N$ (or $\theta_{\rm 1} = 0.9N$ and $\theta_{\rm 2} = N$) for the condition of the LFE (or the HFE). Here $(\theta_{\rm 2} - \theta_{\rm 1})/N$ represents the percentage of the number of initially excited modes in the total. It is expected that $\bar{E}_{\rm in}(t)=1$ when the system enters the fully thermalized state.

As shown in Fig. \ref{fig:EkvsT}(a), the evolution of $\langle\bar{E}_{\rm in}(t)\rangle$ in the diatomic FPUT-$\beta$ chains is independent of $\Delta m$ under either HFE (red lines) or LFE (blue lines). For the LFE, the  $\langle\bar{E}_{\rm in}(t)\rangle$ almost stays in the initial value at the beginning and then begins to decay. While for the HFE, the $\langle\bar{E}_{\rm in}(t)\rangle$ quickly decays to an interim plateau which is little attenuation and then resumes to the fast decay. For both initial conditions, the early attenuation behavior of $\langle\bar{E}_{\rm in}(t)\rangle$ is different, but eventually they both decay to $1$. Besides, the time needed to reach thermalization under HFE is longer. It can be seen from Fig. \ref{fig:EkvsT}(a) that the time difference mainly comes from the duration of metastable state (plateau in the panels). In Fig. \ref{fig:EkvsT}(b), we show the numerical results for the diatomic Toda chains. Clearly, the relaxation behavior of the energy is very different from that of the FPUT-$\beta$ chains. When $\Delta m=0$, the system is reduced to the monatomic Toda chain which is integrable, therefore it is expected that the system cannot be thermalized under any initial conditions. This is confirmed numerically under either HFE (red circles) or LFE (blue circles). For different initial conditions, the $\langle\bar{E}_{\rm in}(t)\rangle$ fast decays to plateaus with different heights. When $\Delta m\ne 0$, either under HFE or LFE, all the $\langle\bar{E}_{\rm in}(t)\rangle$ for various $\Delta m$ first reach the stable value of $\Delta m=0$, and finally reach the value $1$ (thermalization achieved). These results imply that the integrability of the Toda chain is broken by unequal masses. In addition, there exist remarkable differences between the results under HFE and LFE. Firstly, it can be seen from the initial evolution of $\langle\bar{E}_{\rm in}(t)\rangle$ that the energy distributed among the initially excited modes is more effective to transport out under LFE than that under HFE. Secondly, the duration of plateau under HFE is longer than that under LFE. Lastly, after the plateau, the $\langle\bar{E}_{\rm in}(t)\rangle$ under HFE decays sharply to the value $1$ while that under LFE approaches $1$ gently.

\begin{figure}[t]
  \centering
  \includegraphics[width=1\columnwidth]{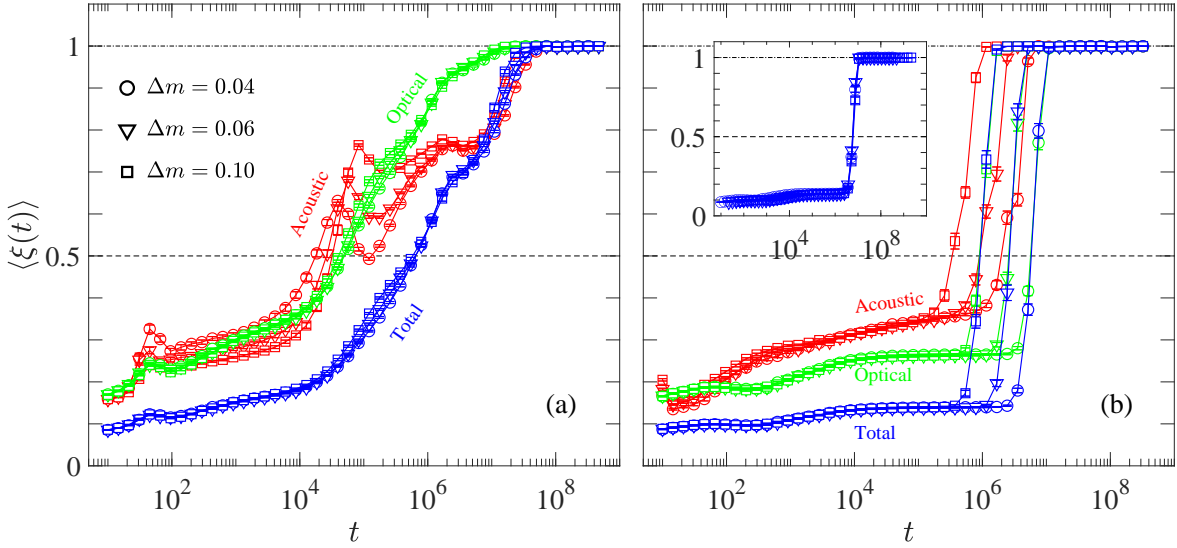}\\
  \caption{(a) The dependence of $\langle\xi(t)\rangle$ on time $t$ for the diatomic FPUT-$\beta$ chain with various mass difference $\Delta m$, in semilogarithmic scale. The red and green data points are the results of defining $\langle\xi(t)\rangle$ on the acoustic and optical branches, respectively. (b) The results for the diatomic Toda chain. Inset: The same as the main panel but the (blue) curves are shifted properly in the horizontal direction (with that for $\Delta m=0.04$ unshifted) so that they perfectly overlap with each other. The number of particles $N=1024$, and the energy density $\varepsilon=0.01$ are kept fixed. }\label{fig:2}
\end{figure}

To observe the thermalization dynamics in detail and to quantitatively characterize the thermalization time $T_{\rm eq}$, we study the time evolution of $\langle \xi(t) \rangle$ defined by Eq.~(\ref{eq:xi}) and the $\langle \xi_{\rm A}(t) \rangle$ and $\langle \xi_{\rm O}(t) \rangle$ which are, respectively, defined on the acoustic modes and the optical ones, namely
\begin{equation}\label{eq:xiA}
  \xi_{\rm A}(t)=M^{-1}e^{\eta(t)},\quad \eta(t)=-\sum_{k=1}^{M}w_k(t)\log[w_k(t)],
\end{equation}
where $w_k(t)=\bar{E}_k(t)/\sum_{j=1}^{M}\bar{E}_j(t)$, and
\begin{equation}\label{eq:xiO}
  \xi_{\rm O}(t)=M^{-1}e^{\eta(t)},\quad \eta(t)=-\sum_{k=M+1}^{N}w_k(t)\log[w_k(t)],
\end{equation}
where $w_k(t)=\bar{E}_k(t)/\sum_{j=M+1}^{N}\bar{E}_j(t)$. Figure.~\ref{fig:2}(a) shows the results of the diatomic FPUT-$\beta$ chains. It is seen that all curves increase from a small value to $1$, and the curves corresponding to various $\Delta m$ nearly overlap with each other, which means that the $T_{\rm eq}$ of the system is independent of $\Delta m$. Note that $\langle \xi_{\rm O}(t) \rangle$ reaches the value 1 first, while $\langle \xi_{\rm A}(t) \rangle$ and $\langle \xi(t) \rangle$ reach simultaneously the value 1 later. Namely, the $T_{\rm eq}$ of the diatomic FPUT-$\beta$ chain is determined by the thermalization of the acoustic modes. As a contrast, Fig.~\ref{fig:2}(b) presents the results for the diatomic Toda chains. It can be seen that all curves stay on a small value for a long time then a jumping behavior occurs, which corresponds to that shown in Fig.~\ref{fig:EkvsT}(b). Notice that $\langle \xi_{\rm A}(t) \rangle$ reaches the value $1$ first, while $\langle \xi_{\rm O}(t) \rangle$ and $\langle \xi(t) \rangle$ reach simultaneously the value $1$ later, which means that the $T_{\rm eq}$ of the diatomic Toda chain is ruled by the thermalization of the optical modes. Besides, note that all the $\langle \xi(t) \rangle$ (blue lines) curves can overlap upon suitable shifts [see the inset in Fig.~\ref{fig:2}(b)], which suggests that taking different values of $\langle \xi(t) \rangle$ to define the thermalization time $T_{\rm eq}$ will not affect its scaling behavior. For both models, we adopt the definition of $T_{\rm eq}$ as that when $\langle \xi(t) \rangle$ reaches the threshold value $0.5$ (as done in Refs.~\cite{Benettin2011,Benettin2013}) to save the cost of calculation since only the scaling behavior but not the specific value of $T_{\rm eq}$ is usually interested.

\begin{figure}[t]
  \centering
  \includegraphics[width=1\columnwidth]{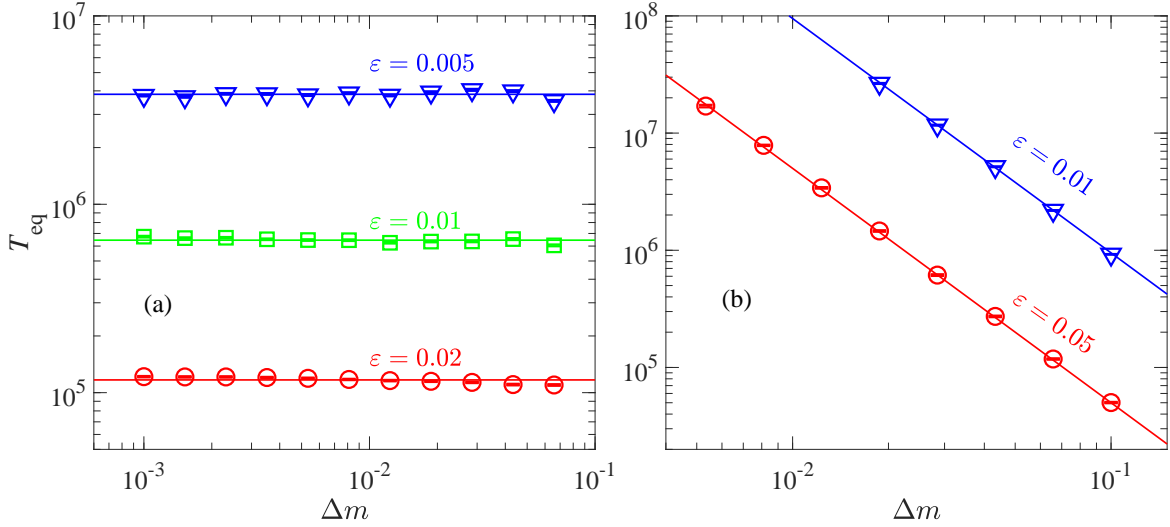}\\
  \caption{(a) The thermalization time $T_{\rm eq}$ as a function of $\Delta m$ for the diatomic FPUT-$\beta$ chain with different energy densities in the log-log scale. The horizontal lines are drawn for reference. (b) The results for the diatomic Toda chains. The solid lines with slope $-2$ are drawn for reference. The number of particles $N=2048$ is kept fixed.}\label{fig:3a}
\end{figure}

In Fig.~\ref{fig:3a}(a), we show the thermalization time $T_{\rm eq}$ as a function of the mass difference $\Delta m$ for the diatomic FPUT-$\beta$ chains with various energy density $\varepsilon$. It can be seen that the $T_{\rm eq}$ is independent of the $\Delta m$ because it does not change the perturbation strength of the system. Figure~\ref{fig:3a}(b) presents the numerical results for the diatomic Toda chains. It can be seen that all the points fall on the lines with a slope of $-2$, suggesting $T_{\rm eq}\propto \Delta m^{-2}$. The results of both models in the HFE are qualitatively identical with those in the LFE (see Fig.~6 in Ref.~\cite{PhysRevE.100.052102}), except that the $T_{\rm eq}$ in the former case is nearly one order of magnitude larger than that in the latter case under the same $\varepsilon$ and $\Delta m$. Note that the $T_{\rm eq}$ sensitively depends on $\varepsilon$ for the two models. The relationship between $T_{\rm eq}$ and $\varepsilon$ is studied in detail below.

\begin{figure}[t]
  \centering
  \includegraphics[width=1\columnwidth]{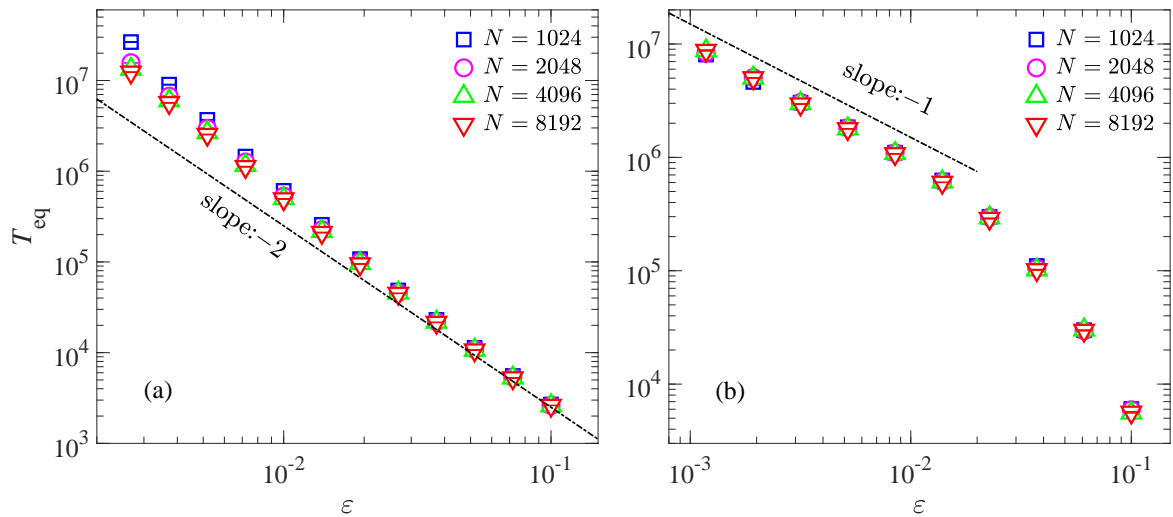}\\
  \caption{(a) The thermalization time $T_{\rm eq}$ as a function of energy density $\varepsilon$, for the diatomic FPUT-$\beta$ chain with different numbers of particles in the log-log scale. (b) The numerical results for the diatomic Toda chain. The mass difference $\Delta m=0.1$ is kept fixed. Energy is initially distributed among $10\%$ of modes of the highest-frequency.}
  \label{fig:TeqVsE}
\end{figure}

Figures~\ref{fig:TeqVsE}(a) shows the dependence of  $T_{\rm eq}$ on $\varepsilon$ at different system sizes for the diatomic FPUT-$\beta$ chains. It can be seen that $T_{\rm eq} \propto \varepsilon^{-2}$, in line with the prediction of the wave turbulence theory \cite{PhysRevLett.120.144301}, though there is a deviation in the region of small $\varepsilon$. Since the overall trend of the deviation decreases with the increase of size, it is considered as the finite-size effect, which has also been observed in previous studies of monatomic chains \cite{Benettin2011,PhysRevE.100.010101}. In Fig.~\ref{fig:TeqVsE}(b), we show the results of the diatomic Toda chains. It can be seen that $T_{\rm eq} \propto \varepsilon^{-1}$ as $\varepsilon$ decreases. For a fixed $\Delta m$, the diatomic Toda chain is considered as a perturbed diatomic harmonic one, and the third-order nonlinearity becomes the leading perturbation with the strength $\sim\varepsilon^{1/2}$ when $\varepsilon$ is small [see again Eqs.~(\ref{eq:TodaExpand}) and (\ref{eq:TodaExpande})]. If the third-order nonlinearity dominates the thermalization behavior of the chain, it is expected that $T_{\rm eq} \propto \varepsilon^{-1}$; i.e., $T_{\rm eq}$ is inversely proportional to the square of the perturbation strength. Compared with the monatomic chains, the optical branch in the diatomic chain increases the scattering space of normal modes, and thus the three-wave process forbidden in the monatomic chain is allowed to occur in the diatomic case, which is supported by the findings that have been reported in Ref.~\cite{pezzi2021threewave} (i.e., the three-wave resonance can occur in the diatomic FPUT-$\alpha$ chain). With the increase of the $\varepsilon$, the third-order nonlinearity gradually loses its dominance, the higher-order nonlinearity holds sway, and thus the $T_{\rm eq}$ decreases as the $\varepsilon$ increases with a steeper slope.

\section{\label{sec:5}Concluding Remarks}

In summary, we have studied the anti-FPUT problem of 1D diatomic chains (i.e., the highest-frequency optical modes are initially excited), by examples of the FPUT-$\beta$ and Toda models. It is shown that the thermalization time $T_{\rm eq}$ of the two models follows the same rule, i.e., $T_{\rm eq}$ is inversely proportional to the square of the perturbation strength. Moreover, there exist some distinctive findings summarized below:

(\emph{I}) \emph{The role of acoustic and optical modes}: For the diatomic FPUT-$\beta$ chains, equipartition within optical modes precedes that within acoustic ones, whereas for the diatomic Toda chain, the order is the opposite. In other words, the acoustic modes in the diatomic PFUT-$\beta$ chains are difficult to thermalize and hence dictate thermalization of the whole system, whereas, in the diatomic Toda chains, the optical ones do. Relaxation under LFE shares these results \cite{PhysRevE.100.052102}.

(\emph{II}) \emph{The lifetime of the metastable state}: Under both initial conditions, the diatomic FPUT-$\beta$ model has no qualitative difference; while the significant difference appears in the diatomic Toda chain. Specifically, under HFE, the metastable state of the diatomic Toda model has a V-shaped energy distribution, while under LFE, the distribution is exponential \cite{PhysRevE.100.052102}. Compared with the results under LFE, the system under HFE has a longer metastable state, and quickly enters the thermalized state once the metastable state starts to destabilize. The mechanism and the process of destablility remain unclear and further study is needed. Due to the longer lifetime of the metastable state, the $T_{\rm eq}$ under HFE is nearly an order of magnitude larger than that under LFE for both models with the same $\varepsilon$ and $\Delta m$.

(\emph{III}) \emph{The choice of the reference integrable system}: The diatomic Toda model presents richer dynamics than the diatomic FPUT-$\beta$ one since the reference integrable system of the former varies in different situations. For example, at a fixed energy density $\varepsilon$, the diatomic Toda chain should be regarded as the Toda chain perturbed by unequal masses thus $T_{\rm eq}\propto \Delta m^{-2}$, while at a fixed $\Delta m$, the system should be regarded as the diatomic harmonic chain perturbed by anharmonicity thus $T_{\rm eq}\propto \varepsilon^{-1}$ for small $\varepsilon$. However, the diatomic FPUT-$\beta$ chain is always the fourth-order nonlinearity perturbation of the harmonic one.

(\emph{IV}) \emph{What's more about diatomic chains}: The diatomic chains have one more optical branch than the monatomic chains, which increases the scattering space of normal modes thus promoting thermalization of the systems. For instance, the three-wave resonance is forbidden in the monatomic chain but it can occur in the diatomic case, which is confirmed by the result of
$T_{\rm eq}\propto \varepsilon^{-1}$ at small $\varepsilon$ for the diatomic Toda chain.

(\emph{V}) \emph{Stubbornness of Toda's integrability}: Although unequal masses destroy the integrability of Toda model, the early stages of thermalization behavior of diatomic Toda chains still show the dynamics characteristics of the monatomic Toda system.

\section*{Acknowledgment}

We acknowledge support by the NSFC (Grants No. 12005156, No. 11975190, No. 11975189, No. 12047501, No. 12064037, No. 11964031, and No. 11764035), and by the Natural Science Foundation of Gansu Province (Grants No. 20JR5RA494, and No. 21JR1RE289), and by the Innovation Fund for Colleges and Universities from Department of Education of Gansu Province (Grant No. 2020B-169), and by the Project of Fu-Xi Scientific Research Innovation Team, Tianshui Normal University (Grant No. FXD2020-02), and by the Education Project of Open Competition for the Best Candidates from Department of Education of Gansu Province, China (Grant No. 2021jyjbgs-06).

\section*{References}

\bibliography{AntiFPUTDiatomic}

\providecommand{\newblock}{}
\begin{thebibliography}{10}
\expandafter\ifx\csname url\endcsname\relax
  \def\url#1{{\tt #1}}\fi
\expandafter\ifx\csname urlprefix\endcsname\relax\def\urlprefix{URL }\fi
\providecommand{\eprint}[2][]{\url{#2}}

\bibitem{Fermi1955}
Fermi E, Pasta P and Ulam S 1955 {\em Los Alamos Scientific Laboratory, Report
  No. LA-1940\/}

\bibitem{dauxois2008PhyToday}
Dauxois T 2008 {\em Phys. Today\/} {\bf 61} 55--57

\bibitem{Campbell2005Chaos}
Campbell D~K, Rosenau P and Zaslavsky G~M 2005 {\em Chaos\/} {\bf 15} 015101

\bibitem{Zabusky2005Chaos}
Zabusky N~J 2005 {\em Chaos\/} {\bf 15} 015102

\bibitem{Zaslavsky2005Chaos}
Zaslavsky G~M 2005 {\em Chaos\/} {\bf 15} 015103

\bibitem{Berman2005Chaos}
Berman G~P and Izrailev F~M 2005 {\em Chaos\/} {\bf 15} 015104

\bibitem{Carati2005Chaos}
Carati A, Galgani L and Giorgilli A 2005 {\em Chaos\/} {\bf 15} 015105

\bibitem{2008LNP728G}
{Gallavotti} G (ed) 2008 {\em {The Fermi-Pasta-Ulam Problem: A Status Report.
  Lecture Notes in Physics}\/} ({\em Berlin Springer Verlag\/} vol 728)

\bibitem{ZabuskySoliton}
Zabusky N~J and Kruskal M~D 1965 {\em Phys. Rev. Lett.\/} {\bf 15}(6) 240--243

\bibitem{PhysRevLett.95.064102}
Flach S, Ivanchenko M~V and Kanakov O~I 2005 {\em Phys. Rev. Lett.\/} {\bf
  95}(6) 064102

\bibitem{FlachPRE2006}
Flach S, Ivanchenko M~V and Kanakov O~I 2006 {\em Phys. Rev. E\/} {\bf 73}(3)
  036618

\bibitem{ZABUSKY1967126}
Zabusky N~J and Deem G~S 1967 {\em J. Comput. Phys.\/} {\bf 2} 126--153

\bibitem{CRETEGNY1998109}
Cretegny T, Dauxois T, Ruffo S and Torcini A 1998 {\em Physica D\/} {\bf 121}
  109--126

\bibitem{Zabusky2006Chaos}
Zabusky N~J, Sun Z and Peng G 2006 {\em Chaos\/} {\bf 16} 013130

\bibitem{dauxois2005anti}
Dauxois T, Khomeriki R, Piazza F and Ruffo S 2005 {\em Chaos\/} {\bf 15} 015110

\bibitem{PhysRevE.61.2471}
Ullmann K, Lichtenberg A~J and Corso G 2000 {\em Phys. Rev. E\/} {\bf 61}(3)
  2471--2477

\bibitem{MIRNOV2001251}
Mirnov V, Lichtenberg A and Guclu H 2001 {\em Physica D\/} {\bf 157} 251--282

\bibitem{Onorato4208}
Onorato M, Vozella L, Proment D and Lvov Y~V 2015 {\em Proc. Natl. Acad. Sci.
  U.S.A.\/} {\bf 112} 4208--4213

\bibitem{PhysRevLett.120.144301}
Lvov Y~V and Onorato M 2018 {\em Phys. Rev. Lett.\/} {\bf 120}(14) 144301

\bibitem{Pistone2018EPL}
Pistone L, Onorato M and Chibbaro S 2018 {\em EPL (Europhysics Letters)\/} {\bf
  121} 44003

\bibitem{PhysRevE.100.010101}
Fu W, Zhang Y and Zhao H 2019 {\em Phys. Rev. E\/} {\bf 100}(1) 010101(R)

\bibitem{Fu_2019Toda}
Fu W, Zhang Y and Zhao H 2019 {\em New J. Phys.\/} {\bf 21} 043009

\bibitem{PhysRevE.94.062104}
Mendl C~B, Lu J and Lukkarinen J 2016 {\em Phys. Rev. E\/} {\bf 94}(6) 062104

\bibitem{PhysRevE.100.052102}
Fu W, Zhang Y and Zhao H 2019 {\em Phys. Rev. E\/} {\bf 100}(5) 052102

\bibitem{Pistone2018}
Pistone L, Chibbaro S, Bustamante M, L'vov Y and Onorato M 2019 {\em Math.
  Eng.\/} {\bf 1}(4) 672

\bibitem{PhysRevLett.124.186401}
Wang Z, Fu W, Zhang Y and Zhao H 2020 {\em Phys. Rev. Lett.\/} {\bf 124}(18)
  186401

\bibitem{pezzi2021threewave}
Pezzi A, Deng G, Lvov Y, Lorenzo M and Onorato M 2021   arXiv:2005.03478

\bibitem{collins1985solitons}
Collins M~A 1985 {\em Phys. Rev. A\/} {\bf 31} 1754

\bibitem{vainchtein2016solitary}
Vainchtein A, Starosvetsky Y, Wright J~D and Perline R 2016 {\em Phys. Rev.
  E\/} {\bf 93} 042210

\bibitem{zhou1996comparative}
Zhou G, Duan Y and Yan J 1996 {\em Phys. Rev. B\/} {\bf 53} 13977

\bibitem{PhysRevE.104.L032104}
Fu W, Zhang Y and Zhao H 2021 {\em Phys. Rev. E\/} {\bf 104}(3) L032104

\bibitem{1967Toda}
Toda M 1967 {\em J. Phys. Soc. Jpn.\/} {\bf 22} 431--436

\bibitem{casati1975stochastic}
Casati G and Ford J 1975 {\em Phys. Rev. A\/} {\bf 12} 1702

\bibitem{PhysRevA.31.1039}
Livi R, Pettini M, Ruffo S, Sparpaglione M and Vulpiani A 1985 {\em Phys. Rev.
  A\/} {\bf 31}(2) 1039--1045

\bibitem{YOSHIDA1990262}
Yoshida H 1990 {\em Phys. Lett. A\/} {\bf 150} 262 -- 268

\bibitem{BIVINS197365}
Bivins R, Metropolis N and Pasta J~R 1973 {\em J. Comput. Phys.\/} {\bf 12} 65
  -- 87

\bibitem{SHOLL1990253}
Sholl D 1990 {\em Phys. Lett. A\/} {\bf 149} 253 -- 257

\bibitem{Benettin2011}
Benettin G and Ponno A 2011 {\em J. Stat. Phys.\/} {\bf 144} 793

\bibitem{Benettin2013}
Benettin G, Christodoulidi H and Ponno A 2013 {\em J. Stat. Phys.\/} {\bf 152}
  195--212

\end{thebibliography}

\end{document}